\documentclass[aps,prl,twocolumn,superscriptaddress,floatfix]{revtex4}

\usepackage{graphicx,graphics}
\usepackage{dcolumn}
\usepackage{amsmath,amssymb,amsfonts}
\usepackage{latexsym,verbatim}
\usepackage{bm}
\usepackage{color}
\usepackage{ulem}
\def\be{\begin{equation}}
\def\ee{\end{equation}}
\def\ber{\begin{eqnarray}}
\def\eer{\end{eqnarray}}

\begin{document}
\title{Graphene field effect transistors as room-temperature Terahertz detectors}
\author{L. Vicarelli}
\affiliation{NEST, Istituto Nanoscienze-CNR and Scuola Normale Superiore, I-56126 Pisa, Italy}
\author{M.S. Vitiello}
\email{miriam.vitiello@sns.it}
\affiliation{NEST, Istituto Nanoscienze-CNR and Scuola Normale Superiore, I-56126 Pisa, Italy}
\author{D. Coquillat}
\affiliation{Universit\'e Montpellier 2, Laboratoire Charles Coulomb UMR 5221, F-34095, Montpellier, France}
\affiliation{CNRS, Laboratoire Charles Coulomb UMR 5221, F-34095, Montpellier, France}
\author{A. Lombardo}
\affiliation{Department of Engineering, Cambridge University, Cambridge, CB3 0FA, UK}	
\author{A.C. Ferrari}
\affiliation{Department of Engineering, Cambridge University, Cambridge, CB3 0FA, UK}
\author{W. Knap}
\affiliation{Universit\'e Montpellier 2, Laboratoire Charles Coulomb UMR 5221, F-34095, Montpellier, France}
\affiliation{CNRS, Laboratoire Charles Coulomb UMR 5221, F-34095, Montpellier, France}
\author{M. Polini}
\affiliation{NEST, Istituto Nanoscienze-CNR and Scuola Normale Superiore, I-56126 Pisa, Italy}
\author{V. Pellegrini}
\affiliation{NEST, Istituto Nanoscienze-CNR and Scuola Normale Superiore, I-56126 Pisa, Italy}
\author{A. Tredicucci}
\affiliation{NEST, Istituto Nanoscienze-CNR and Scuola Normale Superiore, I-56126 Pisa, Italy}
\begin{abstract}
The unique optoelectronic properties of graphene~\cite{review} make it an ideal platform for a variety of photonic applications~\cite{bonaccorso}, including fast photodetectors~\cite{xia}, transparent electrodes~\cite{bae}, optical modulators~\cite{liu}, and ultra-fast lasers~\cite{sun}. Owing to its high carrier mobility, gapless spectrum, and frequency-independent absorption coefficient, it has been recognized as a very promising element for the development of detectors and modulators operating in the Terahertz (THz) region of the electromagnetic spectrum (wavelengths in the hundreds of ${\rm \mu m}$ range), which is still severely lacking in terms of solid-state devices. Here we demonstrate efficient THz detectors based on antenna-coupled graphene field-effect transistors (FETs). These exploit the non-linear FET response to the oscillating radiation field at the gate electrode, with contributions of thermoelectric and photoconductive origin. We demonstrate room temperature (RT) operation at $0.3~{\rm THz}$, with noise equivalent power (NEP) levels $< 30~{\rm nW}/{\rm Hz}^{1/2}$, showing that our devices are well beyond a proof-of-concept phase and can already be used in a realistic setting, enabling large area, fast imaging of macroscopic samples.
\end{abstract}
\maketitle

Photodetection of far-infrared radiation (from hundreds of GHz to a few THz) is technologically important for a wide range of potential applications spanning from medical diagnostics to process control, and homeland security~\cite{mittleman}. THz radiation penetrates through numerous commonly used dielectric materials, otherwise opaque for visible and mid-infrared light. At the same time it allows spectroscopic identification of hazardous substances and compounds through their characteristic molecular fingerprints~\cite{mittleman}. Commercially available THz detectors are based on thermal sensing elements being either very slow 
($10-400~{\rm Hz}$ modulation frequency) (Golay cells, pyroelectric elements), or requiring deep cryogenic cooling ($4~{\rm K}$) (superconducting hot-electron bolometers)~\cite{sizov}, while those exploiting fast non-linear electronics (Schottky diodes) are mostly limited to the range below $1~{\rm THz}$~\cite{sizov}. 

A more recent approach exploits FETs [9], either as III-V high-electron-mobility transistors (HEMTs) or Si-based complementary metal-oxide-semiconductors (CMOSs).  These devices provide excellent sensitivities (NEP in the $10^{-10} - 10^{-11}~{\rm W}/{\rm Hz}^{1/2}$ range depending on temperature and operation frequency), with the intrinsic possibility of high-speed response (in principle just limited by the read-out electronics impedance). This approach has been also demonstrated with self-assembled InAs nanowires, allowing RT THz detection with NEPs of $10^{-9}~{\rm W}/{\rm Hz}^{1/2}$ at $0.3~{\rm THz}$~\cite{vitiello}.

THz detection in FETs is mediated by the excitation of plasma waves in the transistor channel~\cite{dyakonov}. On one hand, a strong resonant photoresponse is predicted in materials having plasma damping rates lower than both the frequency of the incoming radiation and the electron transit time in the transistor channel. This requires electron mobilities of at least a few thousand ${\rm cm}^2/({\rm V} \times {\rm s})$ at frequencies above $1~{\rm THz}$. Under these conditions, stationary states arise from the quantization of plasma waves over the gate length. However, this concept was never demonstrated at room temperature. Furthermore, even though some results have been achieved at cryogenic temperatures~\cite{knap}, thus far no clear-cut evidence was reported for the expected large responsivity enhancement. On the other hand, when plasma oscillations are overdamped, broadband THz detection is observed~\cite{knap}. In this case, the oscillating electric field of the incoming radiation applied between source and gate electrodes produces a modulation of both charge density and carrier drift velocity~\cite{dyakonov}. Carriers traveling towards the drain generate a continuous source-drain voltage $\Delta u$, which is controlled by the carrier density in the channel, and can be then maximized by varying the gate voltage.

A material having high carrier mobility at RT is crucial to take full advantage of the performance enhancement expected from resonant detection, and to make FET detectors equally performant at operating frequencies  
$> 1~{\rm THz}$~\cite{nadar}. The naturally occurring two-dimensional electron gas in a doped graphene sheet has a very high mobility even at RT~\cite{review}. Additionally, it supports plasma waves~\cite{liuplasmon} which are weakly damped~\cite{bostwick,yan} in high-quality samples. Thus, graphene FET plasma-based photodetectors could outperform other THz detection technologies~\cite{ryzhii,rana}.

So far, the development of graphene-based photodetectors has been limited to the near-infrared region, exploiting the creation of electron-hole pairs following light absorption~\cite{xia,lemme,song,gabor}. In the THz range, however, the photon energy is just a few ${\rm meV}$, and light absorption is prevented by Pauli blocking from unavoidable doping of as-produced graphene samples~\cite{casiraghi} or charge inhomogeneities~\cite{martin}. Cryogenic graphene bolometers have also been recently developed and their operation demonstrated in the mid-IR~\cite{yan_MD}.

Here we demonstrate efficient, RT THz detectors, by combining top-gated graphene FETs with THz receiver architectures designed to achieve the desired device sensitivity. When a THz radiation beam is focused on a FET channel, the incoming optical wave is rectified, leading to a dc signal $\Delta u$ between source and drain, proportional to the received optical power~\cite{knap}. To induce a detectable photovoltage signal across the FET, an asymmetry between source and drain contacts is needed, for example in feeding the incoming radiation. This can be achieved either by an asymmetric design of source and drain contacts or by using low shunt-capacitance antennas to funnel the radiation into the strongly sub-wavelength detecting element. Antenna coupling ensures a selective responsivity to both the spatial mode and the polarization of the incoming radiation.

\begin{figure}[t]
\begin{center}
\includegraphics[width=1.00\linewidth]{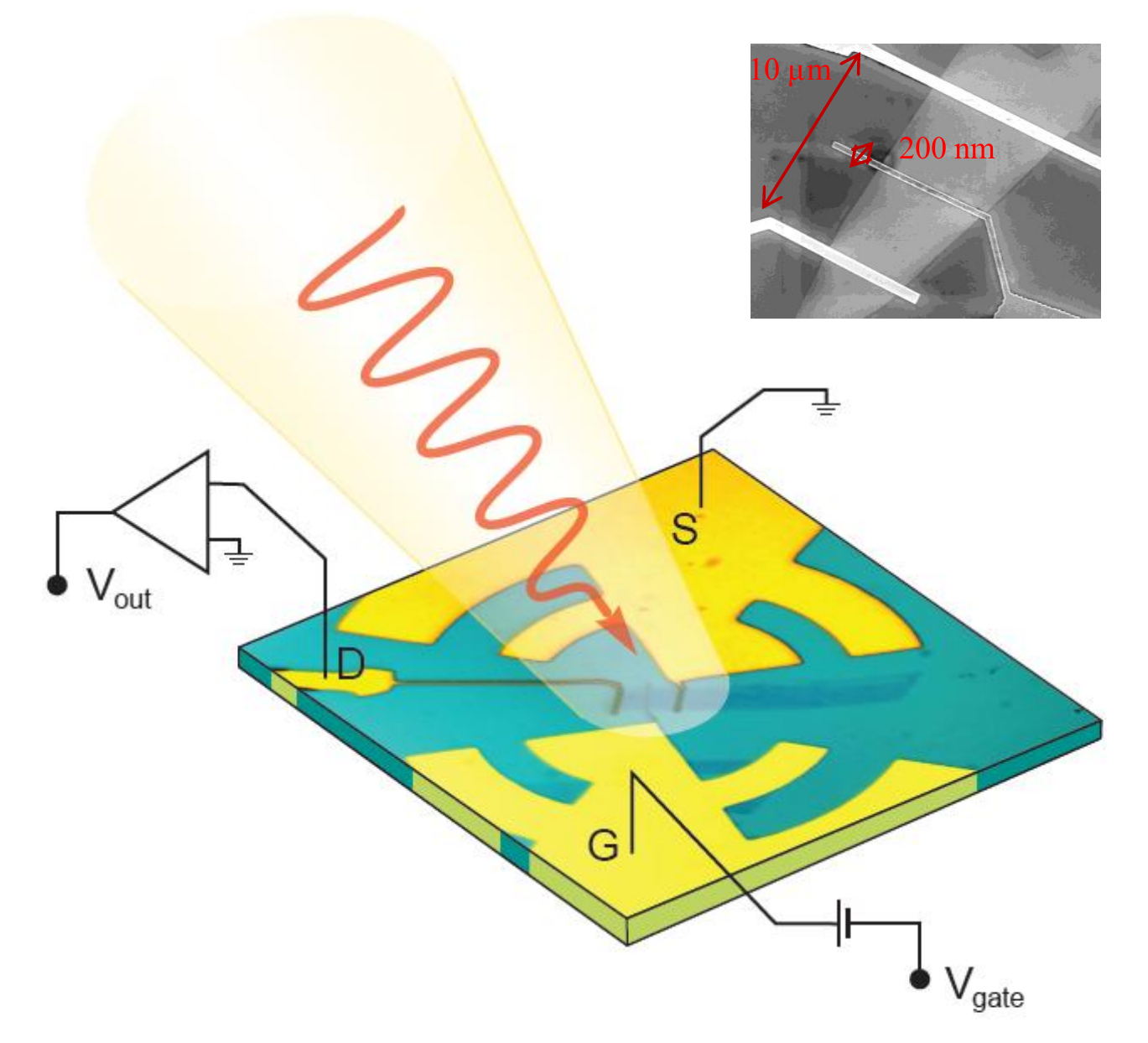}
\caption{(Color online) Schematics of the THz detection configuration in a Field-Effect Transistor (FET) embedding the optical image of the central area of a bilayer graphene-based FET. A log-periodic circular-toothed antenna having an outer radius of $322~{\rm \mu m}$ is patterned between the source and gate contacts, while the drain is a metal line running to the bonding pad. The channel length is $\approx 10~{\rm \mu m}$, while the gate length ($L_{\rm g}$) is $200~{\rm nm}$. The metallic (Cr/Au) top-gate is patterned in the middle of the graphene channel with standard electron-beam lithography and lift-off technique after atomic-layer deposition of a $20~{\rm nm}$-thick ${\rm HfO}_2$ insulating layer. An electromagnetic simulation using a finite-element method (FEM) software (Ansoft HFSS) to calculate the resonance frequencies of the log-periodic antenna (without the bow-tie element) leads to resonant frequencies close to $0.4~{\rm THz}$,  $0.7~{\rm THz}$, $1~{\rm THz}$, and $1.4~{\rm THz}$. The THz radiation at $0.3~{\rm THz}$ is focused on the bow-tie device by off-axis parabolic mirrors with an optical power of $2.1~{\rm mW}$ and a spot size with a diameter of roughly $30$ microns. Inset: Scanning Electron Microscope image of the monolayer graphene FET.\label{fig:one}}
\end{center}
\end{figure}

We use graphene exfoliated on ${\rm Si}/{\rm SiO}_2$ (see Methods) and lithographically define a single lobe of a log-periodic circular-toothed antenna~\cite{vitiello} as source contact, with an outer radius of $322~{\rm \mu m}$, while the drain is a metal line running to the bonding pad. A second identical antenna lobe acting as top-gate contact is then defined by e-beam lithography, after depositing {\it via} atomic layer deposition a $\sim 35~{\rm nm}$ thick ${\rm HfO}_2$ insulating layer. The channel length is between $7$ and $10~{\rm \mu m}$, while the gate length ($L_{\rm g}$) is $200~{\rm nm}$ or ${\rm 300}~{\rm nm}$ depending on the device.  Figure~\ref{fig:one} shows a microscope image of a fabricated graphene FET, embedded within a schematic representation of the THz detection configuration.

\begin{figure}[t]
\begin{center}
\includegraphics[width=1.00\linewidth]{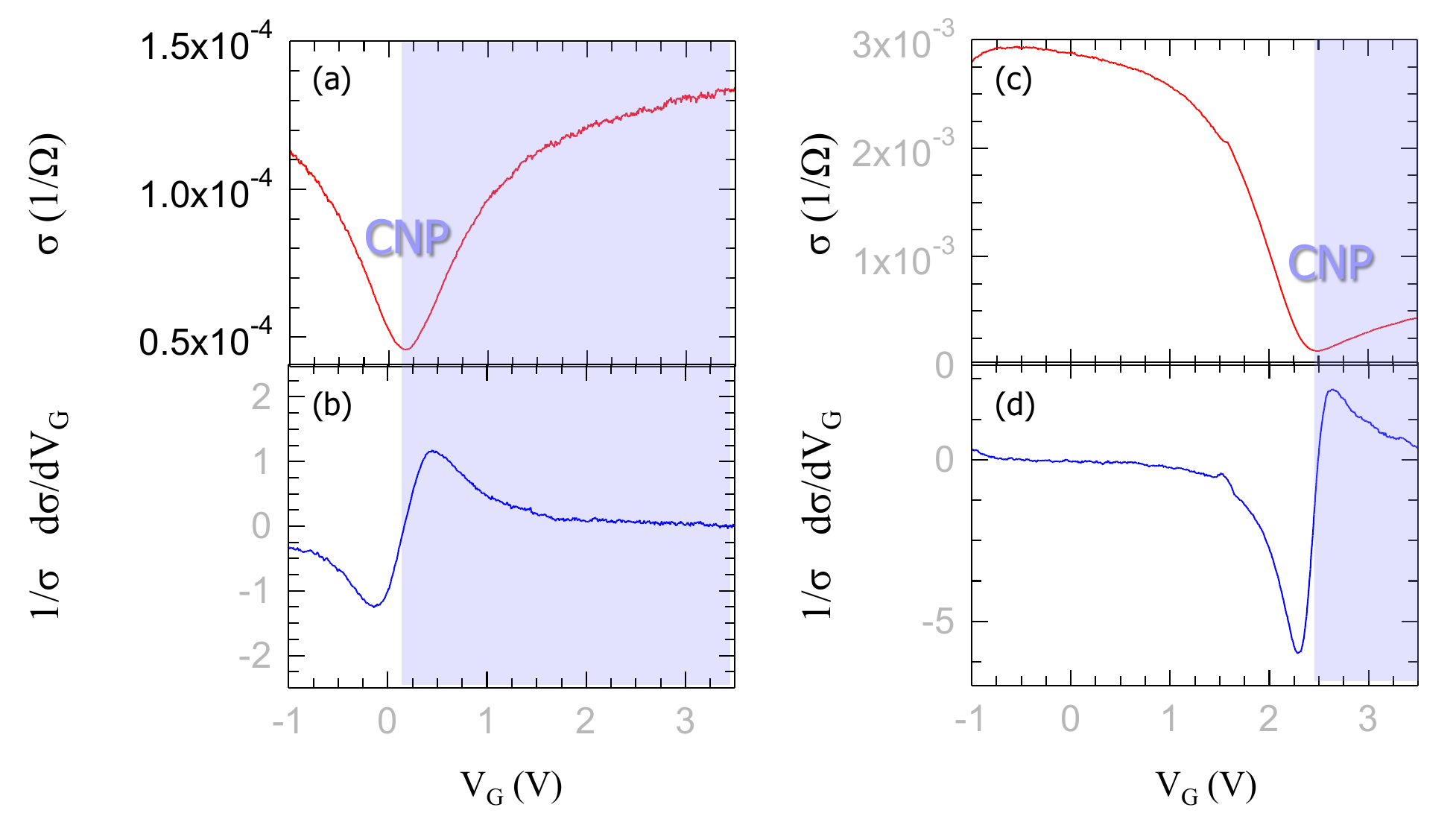}
\caption{(Color online) (a,c) Representative electrical conductivity plots measured at room temperature as a function of the gate voltage in the single-layer graphene (a) and bilayer graphene (c) field-effect transistor while keeping the drain bias at $V_{\rm sd} = 0.1~{\rm mV}$. The drain contact was connected to a current amplifier converting the current into a voltage signal with an amplification factor of $10^4~{\rm V}/{\rm A}$; (b,d) Derivative of the device conductivity as a function of the gate voltage $V_{\rm g}$, plotted once multiplied for the device resistance. This curve represents the expected device responsivity following a diffusive model of transport as described in Ref.~\cite{sakowicz} and in the Methods. The change of sign occurs as $V_{\rm g}$ is swept through the charge neutrality point. \label{fig:two}}
\end{center}
\end{figure}

Figures~\ref{fig:two}(a) and~\ref{fig:two}(c) show the conductivity ($\sigma$) extracted from the current ($I_{\rm sd}$) measured at RT in monolayer (SLG) and bilayer (BLG) graphene devices, respectively, while sweeping the gate voltage ($V_{\rm g}$) from $-1~{\rm V}$ to $+3.5~{\rm V}$ and keeping the source-drain voltage at $V_{\rm sd} = 0.1~{\rm mV}$. In both cases, as expected~\cite{review}, varying the gate voltage, the conductivity goes through a minimum when the chemical potential below the gate is tuned across the charge neutrality point (CNP). In both samples, the CNP is located at positive gate voltages, underlying the p-doping of our graphene samples consistent with the doping level measured by Raman spectroscopy. Note that in our devices the gate electrode length is much smaller than the source-drain distance. This implies that, in general, one cannot simply assume the measured resistance to be fully dominated by that of the gated region.

A diffusive model (Ref.~\cite{sakowicz} and Methods) of transport predicts a second-order non-linear response when an oscillating THz field is applied between gate and source of our FETs. This model yields a photovoltage $\Delta u$ proportional to the derivative of the channel conductivity with respect to the gate bias:
\be\label{eq:deltau}
\Delta u \propto \frac{1}{\sigma}\times \frac{d\sigma}{d V_{\rm g}}~.
\ee
Figures~\ref{fig:two}(b)-(d) show the predicted photovoltage as a function of gate bias for the SLG (\ref{fig:two}b) and BLG (\ref{fig:two}d), as inferred from Eq.~(\ref{eq:deltau}), using as an input the measured gate-voltage-dependent conductivity 
$\sigma =\sigma(V_{\rm g})$. In both cases, $\Delta u$ varies from negative to positive, while crossing the CNP, consistent with the ambipolar nature of charge carrier transport in graphene~\cite{review}.

To access the photoresponse, we shine the $0.3~{\rm THz}$ radiation generated by an electronic source based on frequency multipliers~\cite{tauk}. The radiation is collimated and focused by a set of two f/l off-axis parabolic mirrors. The intensity is mechanically chopped at frequencies varying between $90~{\rm Hz}$ and $1~{\rm kHz}$, and $\Delta u$ is measured by means of a lock-in amplifier in series with a voltage preamplifier having an input impedance of $10~{\rm M\Omega}$ and an amplification factor $G = 1000$. The vertically polarized incoming radiation impinges from the free space onto the graphene devices mounted on a dual-in-line package with an optical power $P_{\rm t} = 2.1~{\rm mW}$.

The photoresponse is measured at zero applied $V_{\rm sd}$, as a dc voltage at the drain, while the source is grounded. The responsivity $R_\nu$ is extracted from the measured $\Delta u$ by using the relation: $R_\nu = \Delta u S_{\rm t}/(P_{\rm t} S_{\rm a})$, where $S_{\rm t}$ is the radiation beam spot area and $S_{\rm a}$ is the active area~\cite{tauk}. This definition of $R_\nu$ assumes that the whole power incident on the antenna is effectively coupled to the graphene FET. In our case, the relatively high device impedance results in a mismatch with the antenna output ($\approx 100~\Omega$ or lower). It is thus likely that a considerable fraction of the radiation field is not funneled into the device. Therefore our $R_\nu$ values have to be considered as lower limit estimations. Given our radiation beam spot diameter $d = 4~{\rm mm}$, we get $S_{\rm t} = \pi d^2/4 = 12.6 \times 10^{-6}~{\rm m}^2$. Since the total area of our device, including antenna and contact pads, is smaller than the diffraction-limited area $S_\lambda = \lambda^2/4$, we take the active area to be $S_\lambda$. The induced $\Delta u$ was then extracted from the measured lock-in signal ${\cal L}$, without any correction from the amplifier input load~\cite{tauk}:
\be
\Delta u = \frac{2\pi \sqrt{2}}{4 G}~{\cal L}~,
\ee
where the factor $2$ is due to the peak-to-peak magnitude, the factor $\sqrt{2}$ originates from the lock-in amplifier rms amplitude, and $\pi/4$ is the fundamental sine wave Fourier component of the square wave produced by the chopper.

\begin{figure}[t]
\begin{center}
\includegraphics[width=1.00\linewidth]{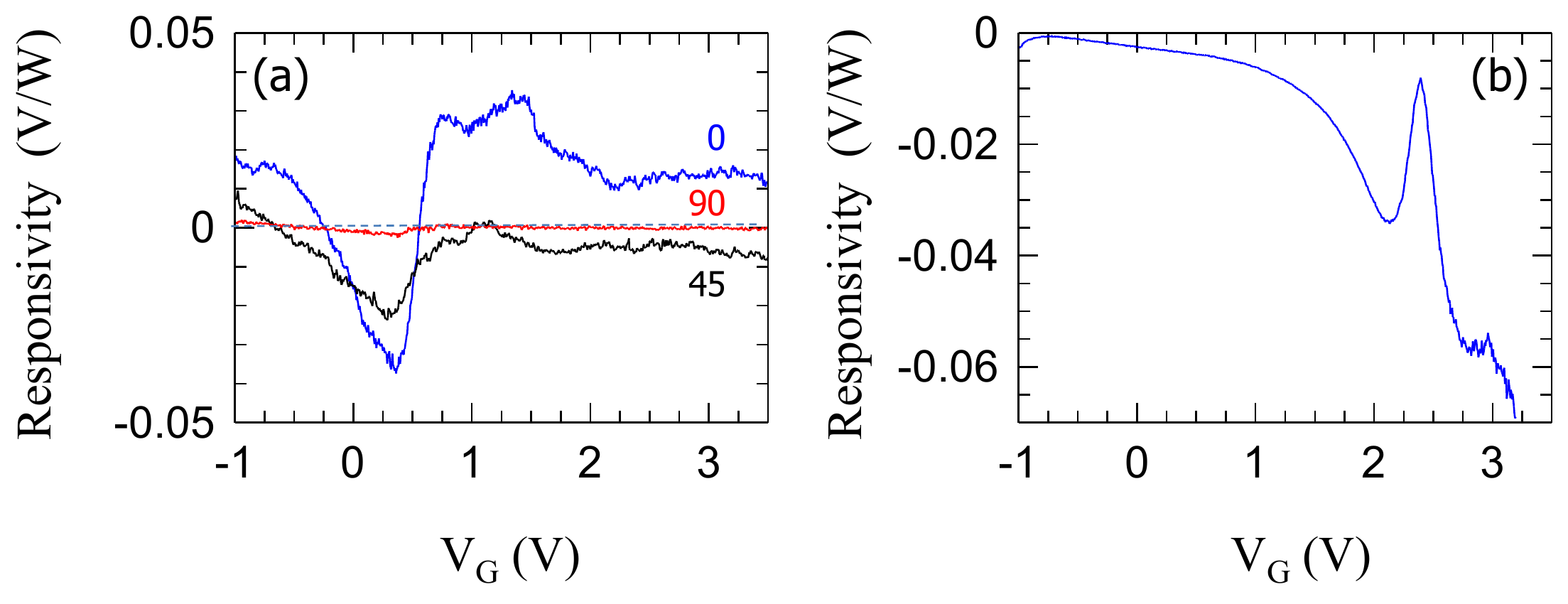}
\caption{(Color online) (a,b) Responsivity values measured as a function of the gate voltage $V_{\rm g}$ at room temperature, for single-layer graphene (a) as function of the angle between the polarization axis of the beam and the antenna axis, and for bilayer graphene (b). The data are extracted from the measured photo-induced voltages across source and drain as described in the text. The incoming $0.3~{\rm THz}$ beam is modulated at $500~{\rm Hz}$ and impinges from the free space onto the graphene devices mounted on the dual-in-line package with an incident optical power $P_{\rm t} = 2.1~{\rm mW}$. \label{fig:three}}
\end{center}
\end{figure}

Fig.~\ref{fig:three}(a) plots $R_\nu$ of the SLG-based FET as a function of $V_{\rm g}$, while modulating the THz source at $500~{\rm Hz}$. Different curves correspond to different relative orientations between the source electric-field polarization and the antenna axis. The photoresponse drops rapidly with angle until it becomes almost zero when the incoming beam polarization is orthogonal to the antenna axis, confirming the efficacy of our dipole antenna. The dependence of $\Delta u$ on $V_{\rm g}$ is in qualitative agreement with the behavior shown in Fig.~\ref{fig:two}(b), thereby proving that in the present conditions our detectors are operating in the broadband overdamped regime. In particular, we note that the sign of the photovoltage changes quite abruptly in the vicinity of the CNP, following the switch of sign of the derivative $d \sigma / d V_{\rm g}$.

We also note that an additional sign switch appears around $V_{\rm g} = 0$ in all curves. This suggests a contribution to the photoresponse of thermoelectric origin~\cite{song,gabor}, arising from the presence of the ungated p-doped graphene regions, and subsequent formation of p-p-p or p-n-p junctions, depending on the gate bias value. It is likely that, owing to the presence of the antenna, the electronic distribution near the junction between gate and source is more effectively ``heated" by the incoming radiation, either through free-carrier (Drude-like) absorption or indirectly from the lattice. The resulting thermoelectric voltage should have the same functional dependence as in Eq.~(\ref{eq:deltau}) but opposite sign, plus a roughly gate-independent value, determined by the thermoelectric Seeback coefficient of the ungated graphene region~\cite{song,gabor}. It should thus change sign near zero gate bias, following the sign of the difference between the thermoelectric coefficients across the junction, as in Ref.~\cite{gabor}.

Our device reaches $R_\nu \sim 100~{\rm mV}/{\rm W}$, when the source is modulated at $1~{\rm kHz}$. This confirms the presence of competing detection mechanisms, each characterized by distinctive response timescales. A quantitative evaluation of each individual contribution is, however, presently difficult, due to their similar functional dependence. As inferred from Eq.~(\ref{eq:deltau}), in the overdamped regime, 
$R_\nu$ is determined by the relative change of conductivity with $V_{\rm g}$; in fact, in this first implementation, our $R_\nu$ is almost one order of magnitude lower than that of nanowire FETs with identical antenna geometries~\cite{vitiello}, but with an on-off current ratio at least one order of magnitude larger.

Figure~\ref{fig:three}(b) plots $R_\nu$ as a function of $V_{\rm g}$ in the BLG-based FET, while modulating the THz source at $500~{\rm Hz}$. The measured shape of the $R_\nu$ curve is in excellent agreement with that predicted by the diffusive plasma-wave detection model [see Fig.~\ref{fig:two}(d)], up to the CNP point. However, no change of sign is measured at the CNP, and a response strongly enhanced with respect to that predicted by Eq.~(\ref{eq:deltau}) appears for $V_{\rm g}$ larger than the CNP, when the device is in the p-n-p configuration.  This suggests an additional contribution to the photovoltage, this time of constant sign. Its magnitude grows rapidly with gate bias, eventually dominating in the regime in which a p-n-p junction is present. We note that the photovoltage is noisier above the CNP, implying an additional intrinsic noise mechanism. We tentatively link this behaviour to interband transitions driven by the THz field at the p-n junction, with the resulting generation-recombination noise. Indeed, this contribution, only seen in BLG, may be related to the higher density of states near the Dirac point, or to the opening of an energy gap due to the electric field, with the THz field possibly adding an energy gap modulation. Whatever the precise origin, this leads to $R_\nu \sim 150~{\rm mV}/{\rm W}$, when the source is modulated at $1~{\rm kHz}$, values even larger than those resulting from the broadband plasma response.
\begin{figure}[t]
\begin{center}
\includegraphics[width=1.00\linewidth]{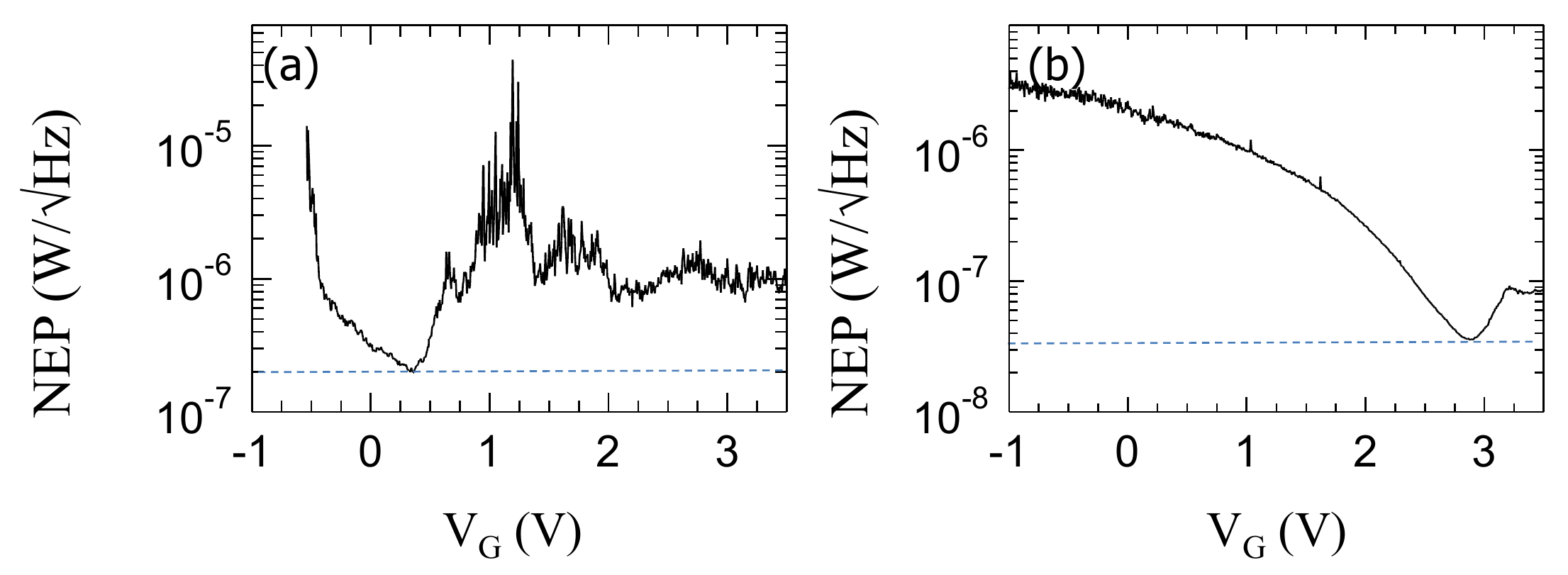}
\caption{(Color online) (a,b) Noise equivalent power (NEP) as a function of the gate voltage measured for single-layer graphene (a) and bilayer graphene (b) assuming that the noise level is dominated by the thermal Johnson-Nyquist contribution 
$N=\sqrt{4k_{\rm B} T/\sigma}$. \label{fig:four}}
\end{center}
\end{figure}

From the application perspective, the relevant figure of merit to characterize a photodetector is the NEP, which corresponds to the lowest detectable power in a $1~{\rm Hz}$ output bandwidth (or equivalently a $0.5~{\rm s}$ integration time). The noise level of FET detectors is dominated by the thermal Johnson-Nyquist contribution~\cite{knap}, expressed as $N=\sqrt{4 k_{\rm B} T/\sigma}$, and can then be extracted directly from the measured conductivity. The NEP is then simply obtained dividing this value by the detector responsivity $R_\nu$. Figures~\ref{fig:four}(a) and~(b) plot our best measured NEPs as functions of $V_{\rm g}$. The minimum RT NEP is $\sim  200~{\rm nW}/{\rm Hz}^{1/2}$ for SLG, and almost one order of magnitude lower ($\sim 30~{\rm nW}/{\rm Hz}^{1/2}$) for BLG. This difference is expected in view of the higher responsivity of the latter, combined with its higher conductivity (translating into lower noise). We note that such numbers are in fact upper limits, and since they refer to the signal power incident on the device as a whole (optical NEP), {\it i.e.} are not corrected for the coupling efficiency of the radiation into the nano-sized transistor element, which is likely very low in our present configuration. Care must be taken when comparing optical NEP values with those derived from the signal power actually absorbed in the detector (electrical NEP), used for example to describe the mid-IR cryogenic bolometers of Ref.~\cite{yan_MD}. Previous RT FET detectors based on InAs nanowires achieved optical NEPs $\sim 10^{-9}~{\rm W}/{\rm Hz}^{1/2}$~\cite{vitiello}, while Si n-MOSs reach $\sim 10^{-11}~{\rm W}/{\rm Hz}^{1/2}$~\cite{knap}. The reason why our devices have higher NEPs stems from the higher $R_\nu$ enabled by the current on-off ratio of the highly mature Si CMOS technology, exceeding $10^5$. We thus expect significant performance improvements to be possible for our graphene FET detectors, even in the overdamped low-frequency regime, for example by reducing the device resistance with shorter channels and raising the mobilities, improving at the same time the coupling with the antenna. We also remark the possibility recently demonstrated to reach on-off current ratios as high as $10^7$ in CVD-grown BLG FETs~\cite{wessely}.

\begin{figure}[t]
\begin{center}
\includegraphics[width=1.00\linewidth]{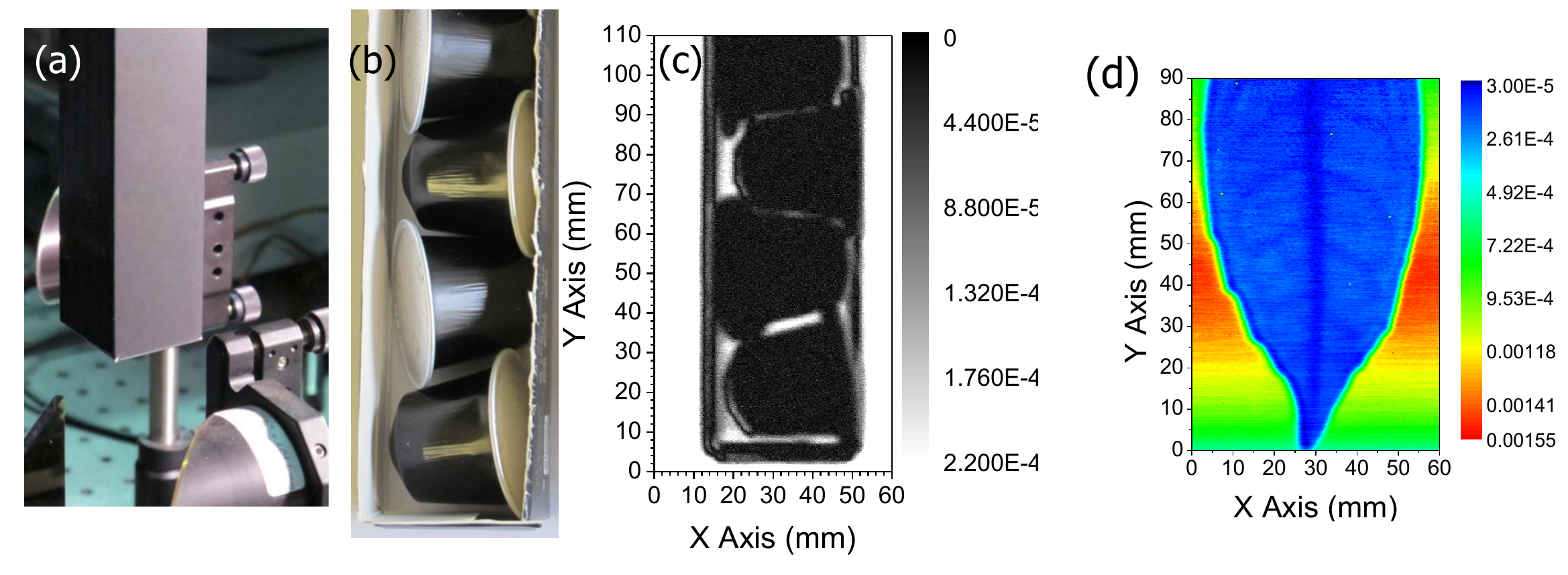}
\caption{(Color online) (a-c) Photograph of a coffee-capsule box (a-b) and $0.3~{\rm GHz}$ transmission mode image of it (b) measured at room temperature with a bilayer-graphene-based FET at $V_{\rm g} = 3~{\rm V}$, mounted on a XY translation stage having a spatial resolution of $0.5~{\rm \mu m}$. The THz image consists of $200 \times 550$ scanned points, collected by raster-scanning the object in the beam focus, with an integration time of $20~{\rm ms}/{\rm point}$; (d) $0.3~{\rm GHz}$ transmission mode image of a leaf measured under the same experimental conditions.\label{fig:five}}
\end{center}
\end{figure}

In any case, our devices are already exploitable, for instance, for large area fast imaging. As test object we used coffee capsules (aluminum foil) inside a cardboard box.  Figure~\ref{fig:five} shows the THz image profile, consisting 
of $200 \times 550$ scanned points, collected by raster-scanning the object in the beam focus, with an integration time of $20~{\rm ms}/{\rm point}$. The coffee capsules as well as the air gaps between them appear clearly in the THz image with a reasonably good spatial resolution. A transmission image of a fresh, green leaf is also shown [see panel (d) of Fig.~\ref{fig:five}].

Our SLG and BLG-based FET detectors can also be used to investigate more fundamental physics. For example, in the weakly-damped limit, they may allow to probe the hydrodynamic behavior of ``chiral" electron plasmas and their non-linear instabilities~\cite{dyakonov,landau}. Moreover, BLG FETs with narrow top gates are particularly suited to unravel the interplay between photo-excited plasma-wave transport and intriguing phenomena such as chirality-assisted electronic cloaking~\cite{gu} and Zener-tunneling-induced negative differential conductivity~\cite{nandkishore}.

\begin{acknowledgements}
We thank Allan H. MacDonald for very fruitful discussions and Vincenzo Piazza for help with the Raman characterization of the samples. Work in Pisa was supported by the Italian Ministry of Education, University, and Research (MIUR) through the programs: ``FIRB - Futuro in Ricerca 2010", Grant no. RBFR10M5BT (``Plasmons and terahertz devices in graphene") and Grant no. RBFR10LULP  (``Fundamental research on Terahertz photonic devices"). Work in Montpellier was supported by the GIS-TERALAB, the GDR2987, the GDR-I THz, and by the Region of Languedoc-Roussillon ``Terahertz Platform". ACF acknowledges funding from ERC grant NANOPOTS, EU grants RODIN and GENIUS, a Royal Society Wolfson Research Merit Award, EPSRC grants EP/GO30480/1 and EP/G042357/1, and the Nokia Research Centre (Cambridge).
\end{acknowledgements}

\appendix

\section{Methods}
\subsection{Fabrication and characterization procedures}

Monolayer (SLG) and bilayer (BLG) graphene flakes were fabricated by mechanical exfoliation from Kish graphite on a highly intrinsic ($\rho = 10~{\rm k\Omega~cm}$) Si substrate with a $300~{\rm nm}$ thermally grown ${\rm SiO}_2$ surface layer. SLG and BLG flakes were identified by optical microscopy and by Raman spectroscopy experiments. Spatially-dependent Raman measurements were carried out exploiting the $488~{\rm nm}$ line of an Argon-ion laser at $1.5~{\rm mW}$. The laser was focused on the sample by a $0.7$-numerical-aperture lens, giving a sub-$\mu m$ illumination spot.

Both samples were then spin-coated with ${\it e}$-beam sensitive resist and contact patterns were exposed by electron beam lithography to realize the drain and source contact pads. ${\rm Cr} (5~{\rm nm})/{\rm Au}(80~{\rm nm})$ Ohmic contacts were then thermally evaporated onto the samples, and lift-off was made in heated acetone. Contact resistivities on the order of $10^{-5}~{\rm \Omega~cm}^2$, were measured. In a successive fabrication step we defined the top gates by ${\it e}$-beam lithography, after preventively depositing {\it via} atomic layer deposition (ALD) a $35~{\rm nm}$ thick ${\rm HfO}_2$ dielectric gate at a temperature $\leq 200~^\circ{\rm C}$. Due to the hydrophobic nature of graphene, a functionalization layer (FL) needs to be deposited before the ALD process. This sticks to graphene and provides the catalytic sites required to start the ALD deposition. A metallic (${\rm Cr}/{\rm Au}$) top-gate channel was then patterned in the middle of the graphene flake. We extracted a top-gate capacitance per unit area of about $300-500~{\rm nF~cm}^2$, assuming a dielectric constant $\kappa = 13-19~{\rm nF~cm}^{-2}$ of the amorphous ${\rm HfO}_2$. Finally Raman spectra measured both before and after the ${\rm HfO}_2$ deposition did not show any evidence of defect-induced D peaks indicating no significant degradation of the graphene properties after oxide deposition. The electrical characterization was performed after having kept the devices under vacuum for $12$ hours. The drain contact was connected to a current amplifier converting the current into a voltage signal with an amplification factor of $10^4~{\rm V}/{\rm A}$.

\subsection{Diffusive transport model}

When the frequency of the impinging radiation $\omega$ satisfies the condition $\omega \tau_{\rm ee} \ll 1$,  $1/\tau_{\rm ee}$ being the electron-electron scattering rate, the electron gas in the channel follows the laws of hydrodynamics~\cite{dyakonov,landau}. Electrons undergo many collisions among each other during one period of oscillation of the external field, thus establishing local thermodynamic equilibrium. The time scale associated with electron-electron collisions in graphene is on the order of $\approx 1- 10~{\rm fs}$ [see, for example, M. Polini {\it et al.}, \prb {\bf 77}, 081411 (2008)]. In the present experiment $\omega/(2\pi)= 0.3~{\rm THz}$, meaning that $\omega \tau_{\rm ee} \approx 2 \times 10^{-3} - 2 \times 10^{-2}$. In this regime, electrons move collectively establishing plasma waves in the channel, which are however strongly damped since $\omega \tau_{\rm tr} \ll 1$. Here $\tau_{\rm tr}$ is the transport time, which can be estimated from the field-effect mobility to be in the tens of ${\rm fs}$. In the hydrodynamic regime, the photoresponse can be calculated by using a one-dimensional model based on Ohm's law,
\be\label{eq:Ohm}
j(x,t) = \sigma E(x,t) = -\sigma\frac{\partial V_{\rm g}(x,t)}{\partial x}~,
\ee
and the continuity equation
\be\label{eq:continuity}
\frac{\partial [-e n(x,t)]}{\partial t} +\frac{\partial j(x,t)}{\partial x} =0~.
\ee
The total carrier density $-en(x,t)$ is modulated by the gate voltage $V_{\rm g}(x,t)$:
\be\label{eq:modulation}
- e n(x,t) = C V_{\rm g}(x,t)~,
\ee
where $C$ is the gate-to-channel capacitance per unit area. The d.c. conductivity $\sigma$ depends on the total carrier density and thus on the gate voltage $V_{\rm g}(x,t)$ by virtue of Eq.~(\ref{eq:modulation}). The system of coupled equations (\ref{eq:Ohm})-(\ref{eq:modulation}) can be solved with suitable boundary conditions, which are imposed at the source ($x = 0$) and at the drain ($x = L_{\rm g}$), with $L_{\rm g} \gg 
c_{\rm s} \tau_{\rm tr}$. Here $c_{\rm s}$ is the acoustic-plasmon group velocity. We now seek solutions of Eqs.~(\ref{eq:Ohm})-(\ref{eq:modulation}) of the form $V_{\rm g}(x,t) = U_0 + U_1(x,t) + U_2(x)$ with the following boundary conditions: $U_1(x = 0,t) = U_a\cos(\omega t)$, $U_1(x = L_{\rm g},t) = 0$, $U_2(x = 0) = 0$, and $U_2(x = L_{\rm g}) = {\rm const}$. This leads to the following source-drain photovoltage:
\be
\Delta u = \frac{U^2_a}{4} \frac{1}{\sigma(U_0)}\left.\frac{d\sigma(V)}{d V}\right|_{V = U_0}~.
\ee

\end{document}